# Information Retrieval in the Cloud

Jochen L. Leidner

***Abstract***

*There has been a recent trend to migrate IT infrastructure into the cloud. In this paper, we discuss the impact of this trend on searching for textual and other data, i.e. the distributed indexing and retrieval of information, from an organizational context.*

## Background

At the time of writing, there is a global corporate trend to reduce the *cost* of company-internal IT infrastructure by giving up self-managed data centers in favor of *cloud computing*, i.e., by embracing cloud-based servers for both storage and computation. The U.S. National Institute for Standards and Technology (NIST) defines cloud computing as "*a model for enabling convenient, on-demand network access to a shared pool of configurable computing resources that can be rapidly provisioned and released with minimal management effort or service provider interaction*" (NIST, cited after BCS (2012)). Amazon, Google, IBM and Microsoft are some examples. At higher levels of the network stack, the level of applications, there is a parallel trend to move away from desktop-installed software packages towards cloud-hosted Web applications, which can reduce the life-cycle cost of installing, maintaining, upgrading and sun-setting applications. SAP, Salesforce, Oracle, ServiceNow and Workday are some examples of major Software-as-a-Service (SaaS) vendors. Both lower-level (storage, computing) and higher-level (application solutions) resources benefit from cloud migrations not just in terms of cost: *security* can be improved by a cloud migration, as cloud vendors can spend more on top-grade security staff in large quantities due to these vendors' scale. *Elasticity* is another huge cloud advantage; the number of servers or application users can be rapidly increased or decreased, since cloud vendor infrastructure is available in abundance (it is shared with all other customers of theirs).

## Managing Corporate Knowledge – Before the Cloud

Before the advent of the clouds, many companies already had heterogeneous networked hardware and software environments in place, and often struggled with data management (i.e., organizing their data assets) and knowledge management (i.e., organizing their knowledge). The transition from self-managed, on-premise servers and company-owned data centers happened gradually, and many organizations went through a phase of relying on internal *private clouds* – sharing resources across departments but only internally. Private clouds have the advantage that all assets remain confined within the organization's walls, legally and physically, which simplifies governance and security. However, private clouds inherit the disadvantages of both worlds: they neither provide the elasticity and scale of public clouds, nor do they benefit from the scale effects of cloud providers:

vendors like Amazon or Google can hire the most expensive security experts, as this cost is distributed across millions of servers.

# Managing Corporate Knowledge – In the Cloud

Private (internal) and public (external) clouds add complexity to these pre-existing challenges in at least three ways: first, access via the network access delays; second, in a cloud scenario data access permeates organizational boundaries, which has security and governance implications. Third, the cloud's elasticity means that rapid commissioning and de-commissioning of cloud storage must be dealt with in terms of incremental index updates to keep search results relevant.

For a comprehensive background introduction into the concepts of cloud computing see Erl, Puttini and Mahmood (2013). Here, we will have to focus somewhat narrowly on the question of *findability* of an organizations in the cloud, which is a large part of effective knowledge management. Anecdotally, about 20% of enterprise employees' time is spend searching for information. Enterprises typically employ Web-based content management systems (CMS), which include enterprise search functions. In a world where enterprises embrace the cloud, a lot of the relevant data that may be needed to meet the needs expressed in a search query will no longer reside inside the (typically centralized) CMS, but scattered across a mix of internal and external (cloud) servers. A new kind of enterprise search is then required to address these needs, and while we cannot address or event attempt a vendor comparison for reasons of space, the table below contains some questions, which most organizations attempting a new implementation of an enterprise search product will have to answer if they want to ensure findability in the organization's cloud.

**Table 1.** Cloud Search: Some Implementation Questions

| No. | Question |
|---|---|
| 1. | What document types are supported by the index crawlers? |
| 2. | Are indexing and retrieval processes federated? |
| 3. | What kind of database and systems connectors are supported by the index crawlers (Oracle RMDB, SAP R/3 ERM, Postgresql/MongoDB/AWS S3, Microsoft Exchange, ...)? |
| 4. | Is the enterprise search architecture aligned with my organization's structure? |
| 5. | What is the average response time for a set of typical queries and under typical system load? |
| 6. | What is the maximum index size (# unique words, # unique documents)? |
| 7. | What is the cost of implementing a particular enterprise search application? How is the cost structured (e.g. by user or by CPU)? |
| 8. | What additional network traffic will implementing a particular solution create on the corporate network? |
| 9. | How is the investment into a solution protected? For example, is there a clause that the source code of the system will be provided if the vendor decides to sun-set the product? |
| 10. | What is the security model (permissions for documents, users, groups)? Does the system support search over encrypted content (e.g. homomorphic encryption)? |
| 11. | How are internal and external cloud resources communicated to the index crawler? Whose responsibility is it to trigger life-cycle state updates, and what API can be used? |
| 12. | What are meaningful and safe default access privileges for indexed cloud data so they can be found using universal enterprise search queries? |
| 13. | What worst-case time guarantees are given in terms of time from storing a new file on an external cloud storage node to that file's content being available in a search? |
| 14. | What policies are available to control index freshness depending on known data volatility (i.e., frequency of change)? |

One of the main risks of cloud use is the improper management of access permissions. Since public cloud resources like Amazon AWS S3 storage buckets reside outside the firewall of the organization, general read access means world-wide read access, so software bugs can have disastrous consequences, from the leaking of trade secrets or to the violation of laws and regulations by disclosing especially protected personal information, while an overly defensive approach leads to cloud silos that cannot be accessed by the organization's cloud search.

## Enterprise Search: Distributed or Federated?

See Baeza-Yates and Ribeiro-Neto (2010) for an introduction to information retrieval, the term of art denoting the sub-domain of information access methods and systems that search technology is part of; it covers indexing and retrieval. White (2015) and Kruschwitz and Hull (2017) provide accessible introductions to enterprise search. Large companies typically have a need for *distributed search*, i.e. there are multiple instances of indexing and retrieval processes operating in parallel (for example, one per country/language pair in which the enterprise operates). The term *federated search* refers to the parallel use of different search engines altogether, not just different instances of the same search engine. While different kinds of data may benefit from deploying specialized search engines, the challenge is to maintain a user experience that hides this complexity by automatically integrating syndicated results. Shokouhi and Si (2011) is a survey of recent research in federated search. Teregowda, Urgaonkar and Giles (2010) provide a digital libraries perspective on cloud search, which in part can generalize to federated search in an enterprise using the cloud.

## Federated Indexing & Retrieval

The published literature on cloud indexing and retrieval is still nascent: Bashir *et al.* (2013) is a review of early research on cloud search. De la Prieta *et al.* (2014) describe a cloud search system intended for educational use, and report small-scale benchmarks. Fang *et al.* (2015) provide an ambitious architecture for semantic cloud search, but without an evaluated implementation. Liu *et al.* (2012) describe methods to efficiently process ranked queries in cloud environments. Pollan and Barreiro (2008) describe gLite, a lightweight distributed environment for cloud search. They also report on a set-up for benchmarking cloud search that relies on the .GOV collection, a well-known Web crawl of public domain World Wide Web pages from U.S. government organizations.

In 2014, the U.S. NIST/DARPA-funded search evaluation workshop TREC convened the most recent instance of the "FedWeb Track", a federated search exercise comprising three tasks, namely (1) vertical selection, (2) resource selection and (3) results merging, and in 2015, a new benchmark collection was made available (Demeester *et al.*, 2015; TREC, 2014; Nguyen *et al.*, 2012).

In the vertical selection sub-task, a set of verticals is determined for a given query. For example, given the search query "definition volatility" and a set of possible vertical indices (finance, news, images, jokes, encyclopedic, travel, ...), the two verticals "encyclopedic" and "finance" could be identified as domains where answers satisfying the information need expressed by the query may likely reside. The resource selection sub-task aims to rank the subset of viable search engines that may retrieve good answers for a query from among the set of search engines that pertain to the chosen verticals. Finally, in the results merging sub-task, the individual returned top-*k* search engine result lists are to be integrated into a single result list ordered from most relevant to least relevant document, given the query. The three-step FedWeb task can serve as a model to understand cloud search, which is also federated. The FedWeb track of TREC does not, however, address evaluating run-time aspects, and the collection is static, very much unlike a real enterprise's data assets, so more work is needed to model and evaluate cloud search in an enterprise (more) realistically. In corporations, cloud servers may either hold their own local indices (Figure 1(a)) or a central index merges the posting lists from all crawled cloud nodes (Figure 1(b)). This idealized picture leaves out practical concerns like replication or sharding of indices for faster access under heavy user load and SaaS applications that may need to be integrated via wrappers that use data export APIs to present the SaaS application as yet another cloud node.

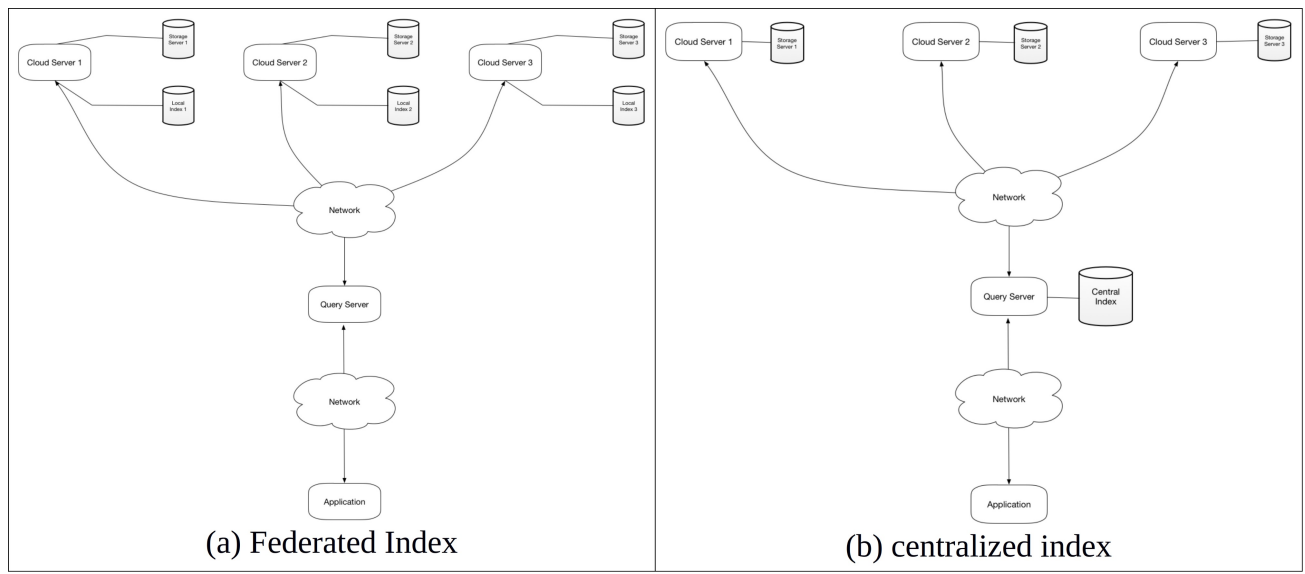

(a) Federated Index          (b) centralized index

**Fig. 1.** Alternative Index Architectures

# Challenges and Opportunities

Besides search, interactive exploration needs to be cloud-enabled; for instance, Baille, Crestani and Carman (2011) describe a federated version of probabilistic topic models. The SaaS paradigm means that increasingly, software functionality, not just data, is available in the cloud, and eventually the search functionality should include a search for software (Dikaiakos, Katsifodimos and Pallis, 2012). Timely access in the limit means real-time search in the cloud, the immediate availability of data in search results with hard guarantees (Uddin *et al.*, 2013). There is an opportunity for the IR research community to develop shared tasks related to cloud search, for instance as part of the TREC (Voorhees and Harman, 2005) or CLEF (Jones *et al.*, 2017) initiatives. For safety-critical information, it may be desirable to search for it without decrypting it, which becomes possible using *homomorphic encryption* (Anand and Satapathy, 2016).

## Summary & Conclusion

In this paper, we have opened the discussion what the migration to the cloud means for search in an organization's environment. We may conclude that more research into federated computing, in particular federated indexing, retrieval, caching and replication is required. A challenge will be to strike the balance between maintaining index completeness across local and cloud resources while respecting access permissions and providing a fast and effective search experience for the users.